\documentclass[sigconf]{acmart}
\usepackage{array}
\usepackage{rotating}
\usepackage{multirow, makecell}
\usepackage{wrapfig}
\usepackage{pdfpages}
\AtBeginDocument{%
  }

\copyrightyear{2025} \acmYear{2025} \setcopyright{rightsretained} \acmConference[CHI PLAY Companion '25]{Companion Proceedings Annual Symposium on Computer-Human Interaction in Play}{October 13--16, 2025}{Pittsburgh, PA, USA} \acmBooktitle{Companion Proceedings Annual Symposium on Computer-Human Interaction in Play (CHI PLAY Companion '25), October 13--16, 2025, Pittsburgh, PA, USA} \acmDOI{10.1145/3744736.3749360} \acmISBN{979-8-4007-2023-9/2025/10}

\begin{document}

\title[Designing Doable and Locally-adapted Action Cards for Bottom-Up Flood Resilience]{Designing Doable and Locally-adapted Action Cards for an Interactive Tabletop Game To Support Bottom-Up Flood Resilience}

\author{Linda Hirsch}
\affiliation{%
  \institution{University of California Santa Cruz}
  \city{Santa Cruz}
  \country{USA}}
\email{uxresearch@hirschlinda.com}
\orcid{https://orcid.org/0000-0001-7239-7084}

\author{James Fey}
\affiliation{%
  \institution{University of California Santa Cruz}
  \city{Santa Cruz}
  \country{USA}}
\email{jfey@ucsc.edu}
\orcid{0000-0002-5033-0134}

\author{Katherine Isbister}
\email{kisbiste@ucsc.edu}
\affiliation{
  \institution{University of California Santa Cruz}
  \streetaddress{1156 High St}
  \city{Santa Cruz}
  \state{California}
  \country{USA}
  \postcode{95064}
}\orcid{0000-0003-2459-4045}

\renewcommand{\shortauthors}{Hirsch et al.}

\begin{abstract}
Serious games can support communities in becoming more flood resilient. However, the process of identifying and integrating locally relevant and doable actions into gameplay is complex and underresearched. We approached the challenge by collaborating with a community-led education center and applying an iterative and participatory design process of identifying and defining actions that may increase local applicability and relevance. The process comprised a field observation, two expert focus groups (n=4), and an online survey (n=13). Our findings identified 27 actions related to increasing or maintaining individuals' and communities' flood resilience, which we turned into 20 playing cards. These action cards are a part of a larger interactive tabletop game, which we are currently developing. Our work discusses the potential of card games to educate non-experts to increase flood resilience, and contributes to our process of identifying local needs and conditions, and turning them into engaging game artifacts for bottom-up empowerment.
\end{abstract}

\ccsdesc[500]{Human-centered computing~Collaborative and social computing systems and tools, bottom-up}
\ccsdesc[300]{Human-centered computing~Interactive systems and tools}

\keywords{serious game, climate, resilience, flood, card game, interactive tabletop, survey}

\begin{teaserfigure}
  \includegraphics[width=\textwidth]{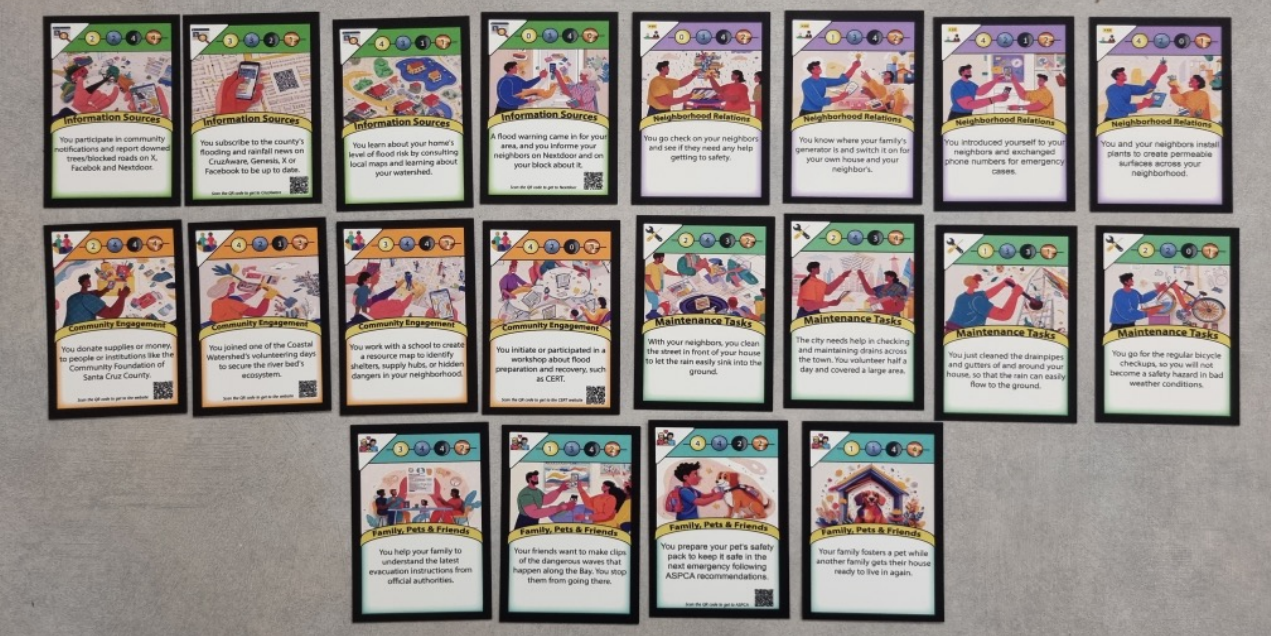}
  \caption{These 20 cards communicate doable actions and local information sources to better prepare for, deal with, or recover from flooding scenarios. }
  \Description{The teaser figure shows the 20 action cards that we developed through our approach. Actions relate to, e.g., subscribing to the local alert system or checking on your neighbors.}
  \label{fig:teaser}
\end{teaserfigure}

\maketitle
\section{Introduction}
Mitigating natural hazard risks, such as floods or wildfires, and adapting to their impacts is essential for individuals and communities living in risk areas. It requires individuals to have a sense of agency and understanding of their role~\cite{Brown2011,Hirsch2025} and impact in adapting to and resisting against a natural hazard~\cite{sharma_whose_2024,Johns2025}. However, available tools and measures shared by official authorities are often too abstract, complex, and hard to apply on a local community level~\cite{Brown2018,Krishnan2021,Vivekananda19052014,Galeote2024}, preventing locals from implementing countermeasures and increasing their resilience. Individuals often feel they have too little agency to impact the course of a natural hazard and implement countermeasures, or do not understand the urgency and need to take precautions~\cite{paton2019risk,Hirsch2025}. These challenges emphasize the lack of awareness, understanding, and agency to take suitable and timely counter- or adaptation measures against natural hazard risks. 

Serious games have been shown to increase awareness and understanding of the personal relevance and impact of topics such as climate change~\cite{Galeote2024, Galeote2023,Troiano2020}, wildfire~\cite{Johns2024,Johns2025}, or flooding~\cite{TAILLANDIER2025}, particularly when aiming to engage local communities~\cite{Maram2024}. Playful approaches further facilitate community engagement, and ongoing work has focused on community-driven initiatives and bottom-up empowerment and education~\cite{Rigby2023_climate, Jensen2024, Johns2024}. One of the strategies suggests the integration of local conditions and resources, which can lead to individuals' better understanding and adaptability of solutions, while emphasizing their personal relevance~\cite{choko2019,ensor2018,Marzouki2018,Bove2022}. Furthermore, integrating actions doable for non-experts to better prepare, deal with, or recover from a natural hazard can increase the sense of agency and empowerment~\cite{Murphy2015,sharma_whose_2024}. However, such doable, locally-adapted and playful interventions are complex to develop and are, thus, still rare~\cite{Gardiner2019}. Similarly, approaches leading to identifying relevant actions and transferring them into game mechanics are underexplored~\cite{green2024}. 

We approached this gap by taking an iterative and participatory design approach, comprising a set of field observations, two focus groups with experts, and one online survey, aiming to identify locally-adapted actions that non-experts can do to increase their own, their families, and their community's flood resilience. We focused on flood resilience as the region in which we conducted the research, Santa Cruz, California, is prone to coastal and river flooding and precipitation. This work is part of a larger project that aims to develop an informal educational game for and with a community-led educational center, the Seymour Center. Here, we focus on the process of identifying the actions and their integration into the overarching game concept in the form of playing cards, see \autoref{fig:teaser}. 

Our results identified 27 locally-adapted actions doable for locals that we turned into 20 playing cards as part of a larger interactive tabletop game, which is currently being tested at the time of submission. 
Our work contributes to developing local flood resilience from the bottom up through playful, intergenerational experiences. We see value in sharing our approach of identifying and transferring locally relevant actions into game components, as it has been a continuous challenge to develop such interventions~\cite{Gardiner2019, Rigby2023,green2024}. We further contribute the resulting game artifact, which will be evaluated on how card games co-developed with the local community can serve the local community. We see both our approach and the exploration of card games for climate resilience as relevant contributions for other game developers, designers, and researchers that we would like to discuss further with the CHI Play community.

. 


\section{Related Work}

\begin{figure*}
    \centering
    \includegraphics[width=\textwidth]{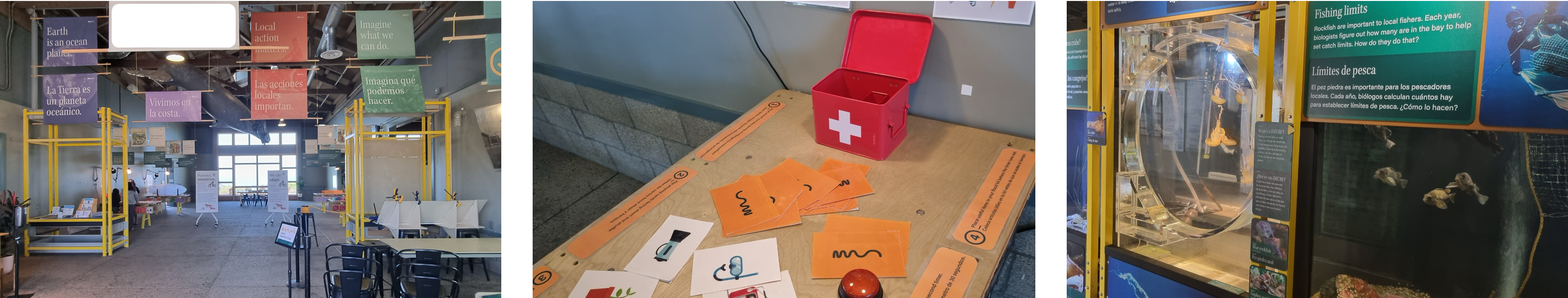}
    \caption{Impressions at the center: left) The entry hall shows signs in English and Spanish, a seating area, and tables with installations. Middle) cards to pack an emergency kit, which did not provide any feedback on visitors' choices. Right) Aquariums with sea creatures.}
    \label{fig:seymourVisit}
    \Description{The figure consists of three images, showing impressions of the center. At the very left, you can see the entrance area with a large hall, information signs, and different tables and seating areas. In the middle, there is a photo of one of the existing non-interactive playful interventions. The intervention teaches about what people should pack for their emergency kit using printed cards. The picture on the right shows two of the aquariums that are part of the center's exhibit.}
\end{figure*}

\subsection{Communities' and Individuals' Flood Resilience}
Flood resilience describes the ability to prepare for, deal with, and recover from flood-related disruptions and impacts~\cite{ALABBAD2021,Auliagisni2022}. Increasing flood resilience requires considering and connecting a variety of resources, skills, and local conditions, including impacts on infrastructural, social, environmental, and organizational levels~\cite{PRASHAR2023, JACINTO2020}. Communities with strong social neighborhoods and socially responsible individuals prove to be more resilient than other communities~\cite{Mullins2011,Soetanto2017,ROBERTSON20211,Suhaimi2022,Andrew2012,Patel2017}, because those members have a more distinct sense of shared responsibility and openness to change~\cite{ROBERTSON20211}. For increasing the sense of social responsibility in community members, \citet{markantoni_can_2019} emphasises the need for successful engagement experiences in the form of doable shared projects. In line with their findings, \citet{Davies2015}'s work showed that the sense of control is crucial for motivating individuals and giving them a sense of empowerment. Another contributing factor for individuals' and communities' resilience is their education about local flooding risks and countermeasures that are doable and relevant for them~\cite{Munsaka2023,Auliagisni2022}. However, methods that focus on bottom-up education and empowerment toward flood-resilience are scarce as they are complex to develop with limited transferability~\cite{Gardiner2019,Davies2015}.

\subsection{Climate-Related Playful Interventions and Games}
Serious games and playful concepts have been shown to increase engagement and facilitate understanding of complex topics~\cite{Bogdan_2024,Troiano2020,Maram2024,Detten2024,Galeote2023_game,green2024}, such as climate change and natural hazard impacts. \citet{Kadir2024} explored a flood simulation game using Unity to educate players about pre-flood mitigation possibilities and evacuation behaviors. Similarly, \citet{Bogdan_2024} developed a serious digital game for engineers to understand the complex social and cultural impact that engineering solutions to reduce flood impacts may have on society. Another work by \citet{TAILLANDIER2025} presents the benefits of dynamic sketch maps to provide easy-to-understand and clear information relevant for locals. Their work further emphasizes the need and benefit of triggering discussions among players for reflection and integrating locals' perspectives into the gameplay. While researchers did not observe differences in the learning effect between traditional text-based versus video or virtual reality game formats~\cite{Galeote2023,Galeote2023_game}, familiarity and familiar features integrated into the games can facilitate learning~\cite{Gagne1985,Galeote2024}. However, methodological approaches of identifying and transferring local specifics and dependencies into gameplay are underdeveloped and underresearched~\cite{green2024,Galeote2023}.

\subsection{Card Games for Community Engagement}
Cards as a platform allow for the hosting of information through text, icons, color, and other forms of art and design to deliver information about the game and real-world topics~\cite{altice_playing_2014,turkay_collectible_2012}. In a discussion of collectible card games (CCGs), \citet{turkay_collectible_2012} list three factors of CCGs that aid in learning. 
The social nature of CCGs provides opportunities for apprentice models to develop between players, and discussions of strategy can lead to more engagement with the educational backing of serious games. 
When applied to community-focused efforts, card games can be empowering to the members of the community as they provide an understandable structure to complex real-world issues in a way that allows for those without domain-specific expertise to engage in meaningful ways~\cite{menconi_card_2020,mccurry_its_2025}. In both \cite{altice_playing_2014,turkay_collectible_2012}, the CCGs were designed with expert and community involvement to increase understandability.

\section{Methods \& Results}
We developed the action cards as part of an informal serious tabletop game for the center through an iterative and participatory design approach. We will present each step and its results by method for better readability.

\subsection{Overall Approach}
Our approach comprised an onsite field observation at the center, two focus groups (n=4 experts each), and a small online survey (n=13 locals). We first spent two hours observing and taking notes at the center, which would give us a more objective understanding of the community-led communication and education. This was followed by the first focus group with center experts and an online survey (n=13) to gather information about what locals already do and the reasons preventing them from engaging (see \autoref{qualtricsSurvey}). Based on these results and additional online research, the author team created a first set of doable and locally-adapted activities that were then discussed in a second focus group, resulting in 20 action cards. The study received approval from the authors' institution's ethics board.

\subsection{Onsite Field Visit and Early Concept Development}
As this work is part of creating a larger game for the center, we conducted a field observation to learn about onsite conditions, including the type of visitors and other installations. For this, one author took notes and pictures for about two hours at the center. We purposefully entered the field observation with as little prior knowledge as possible to enable an open-minded and naive observation of the onsite situation, as known from phenomenological approaches~\cite{Wertz2005,Overgaard2015}.

\begin{figure*}
    \centering
    \includegraphics[width=0.8\textwidth]{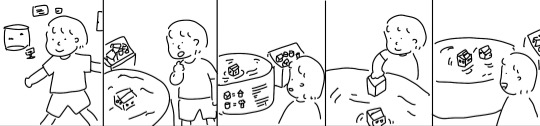}
    \caption{Interactive tabletop storyboard: We used the online collaboration tool \textit{Miro} to brainstorm and create storyboards for our initial concepts. The sketch shows how a child can interact with tangibles on a table to trigger different flooding scenarios.}
    \label{fig:storyboardExample}
    \Description{The figure shows part of the storyboard that we used at the beginning of our project to discuss our tabletop game idea with experts. It shows how a child is coming up to a table and moves stuff around to see how it might change the flooding scenario.}
\end{figure*}

\paragraph{Results} Our observations showed that the center is split into three zones, one for gathering and sitting together, one for educational installations, and one for aquariums. \autoref{fig:seymourVisit} shows impressions from each zones. The center was mainly visited by families (adults and children, seemingly under 12) at the time of observation. Educational installations were non-technological; some were more text-heavy, some more playful, but without providing any feedback or reaction. All installations combined tangible artifacts and textual information in English and Spanish and were set up on tables. We used these insights to brainstorm concepts among the authors as preparation for the first focus group, including, e.g., an interactive tabletop game combining tangible interaction and projected digital output as shown in \autoref{fig:storyboardExample}.

\subsection{Focus Groups}
We conducted two focus groups, one before and one after the online survey. Both included the same group of four experts from the center. Focus groups provide structured discussions to identify needs, values, and priorities relevant for an interaction design concept~\cite{goodman2012observing}. By involving these experts, we had access to local knowledge about flooding risks and sensible countermeasures led by official authorities and institutions. Also, the center is a community-led and educational meeting point for the broader public, emphasizing the experts' knowledge and integration into the local social and cultural values. We further decided to engage experts because their insights are known to be highly efficient in early research phases~\cite{bogner2009interviewing}. Both focus groups took about an hour and were held at the center to reduce the experts' working time. All gave their informed consent. Two researchers took notes at all times. 

\subsubsection{Focus Group I} The first focus group was split into two parts: the first half aimed at understanding what current needs, challenges, and gaps are to having a flood resilient community, and how to empower non-experts to become resilient. For this, we had a semi-structured discussion with questions prepared by us, including \textit{``What are the biggest gaps and challenges between what you as experts know and do and what locals (aka. individual non-experts) know and do?''},  \textit{``What are countermeasures that locals can do to be more flood resilient?''}, \textit{``What are countermeasures that locals can do on a personal versus a communal level?''}, and \textit{``Based on your experience, what are good ways of engaging the local public?''}. In the second half, we discussed early interaction concepts using storyboards (see \autoref{fig:storyboardExample}). We will not go into further details of the other storyboards, as these are not relevant for the card game development.

\paragraph{Results} The expert confirmed the broader public's knowledge and awareness gap regarding flooding risks and countermeasures. They shared links to official websites with further information targeted at the broader public, such as flood safety tips provided by FEMA\footnote{FEMA is the abbreviation for the Federal Emergency Management Agency based in the United States.}. Otherwise, they were unaware of locally-adapted actions that are easily doable and relevant for non-experts. Nonetheless, the discussion led to aligned decisions for the game. For one, experts required that it emphasizes the individual's impact and co-dependencies on their community's flood resilience. This further points toward a collaborative interaction experience that shows how individuals' actions might impact a community's flood resilience. Also, the onsite setup should fit the existing installations, leading to the decision to focus on an interactive tabletop game with tangibles for visitors to interact with. In addition, the center is visited mainly by families, including children from the age of 4 up to the elderly over 65, making it important to have the installation be engaging for all age groups. One expert expressed the wish to connect the game to other local places and institutions dealing with flooding risks and go beyond the onsite experience. The focus group results shaped our subsequent design process and triggered the decision to distribute an online survey to the broader public to learn about flood-related preparation and countermeasures that people already take. 

\paragraph{Focus Group II}
The second focus group aimed at revisiting and discussing the identified actions from the online survey and online research, and how they would best fit into the overarching game concept. For this, we prepared a list of the actions, five action categories, and a suggestion of when we would consider each action most suitable to be done in relation to a flooding event, either long before, briefly before, during, or after. We first gave an update on the game concept and the action list before splitting the group into pairs to review and iterate on the list. The task was to produce a comprehensive list of at least four actions per category, appropriately phrased for the center's target audience. Additionally, we asked the experts to score each activity for each flooding phase from ``zero: not at all'' to ``four: highly relevant'' so that we could transfer the scores into game points.

\begin{figure}
  \begin{center}
    \includegraphics[width=0.5\textwidth]{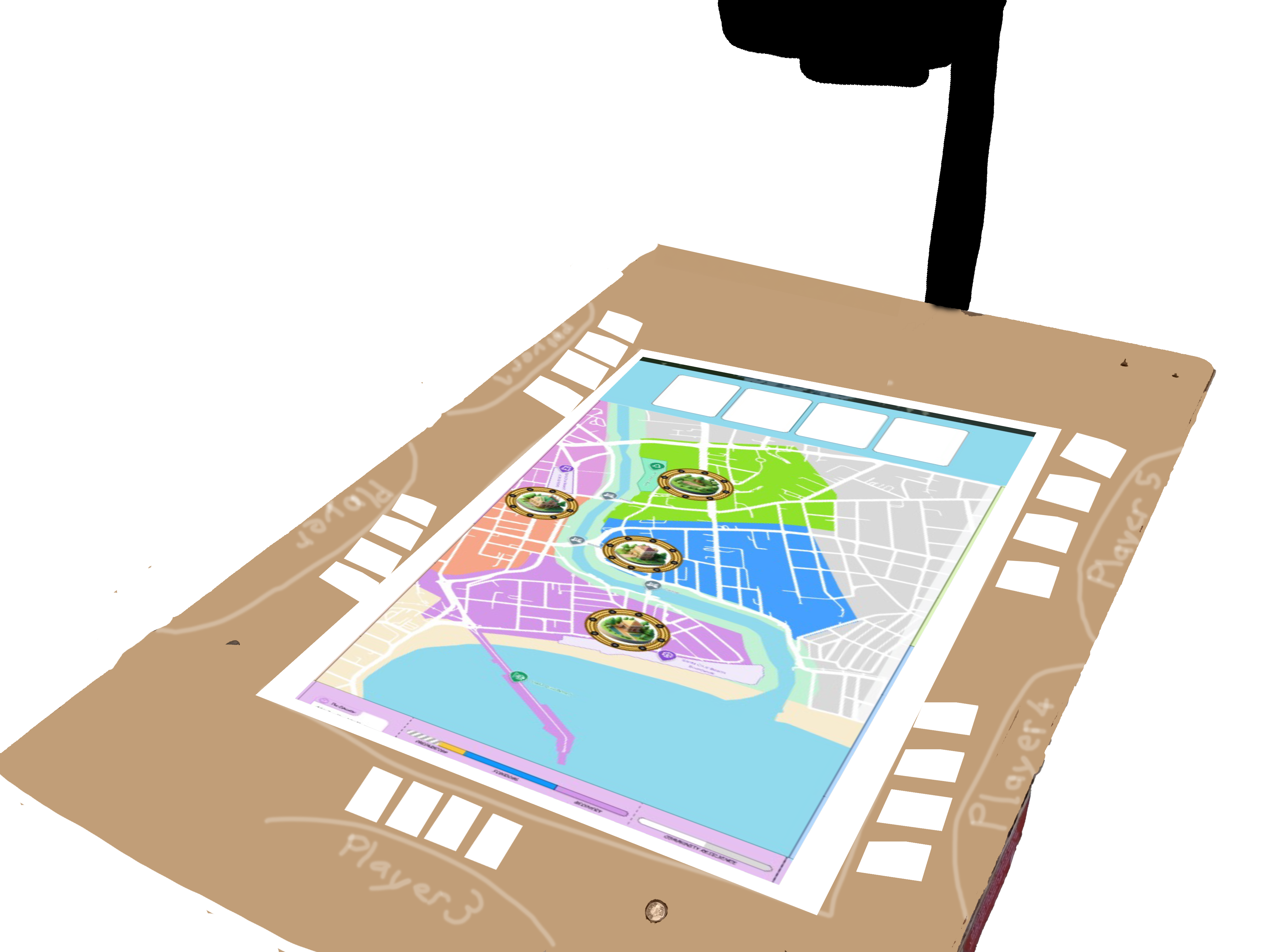}
  \end{center}
  \caption{Mockup of the tabletop and card game setup.}
  \label{fig:tabletop}
  \Description{This figure shows an early mockup of the complete game setup and how the cards are part of the tabletop game. You see a projected map in the middle of the table and five player stations, each with four cards in front of them.}
\end{figure}

\paragraph{Results} The second focus group resulted in a list of actions (see \autoref{tab:actionItems} in the attachment) doable and appropriately phrased for non-expert center visitors. The limitation was raised that younger children will not be able to read. However, these children are normally under the supervision of an adult who can read and show the cards to the children. Yet, this emphasised the need to provide alternative means for the younger children to be engaged and entertained. This led to the design decision of working with flood simulations with integrated and kid-friendly animated characters. \autoref{fig:tabletop} presents a mockup of the installation, including a projector at the top displaying a map of the local high-flood risk zones and the animated characters, as well as the action cards distributed around the table. The setup is future work and serves only as an explanation of how the cards are integrated into the larger game concept.

\subsection{Online Survey and Action Identification}
Between focus groups, we conducted an online survey to gather insights from locals about their countermeasures for becoming more resilient to coastal climate risks, including flooding, and the reasons that prevent them from taking action. We created the survey in Qualtrics and distributed it across the university campus and on bus stops via posters, local Facebook groups, and via the center's email network. In spite of this effort, we received a limited number of 13 responses (11 self-identified as female, 1 as male, 1 preferred not to say). The survey included questions about what participants have done to prepare or deal with potential coastal climate risks, what children and the elderly can specifically do, and what other activities they can think of in this context (see \autoref{qualtricsSurvey} for details). All participants provided digital consent. Survey completion took about $M$=14 min.

\paragraph{Results} Participants' ages ranged from 19 to 77 years old, $M=38$. Four were students, three worked in the education sector, two were retired, three worked as general managers, and one worked as a consultant. Five sometimes participated in climate resilience-related activities, five rarely, two often, and one never. In total, we identified about 27 action items from participants' suggestions. These actions included home and neighborhood hardening activities, subscription to relevant social media channels, and the sharing of information with neighbors, friends, and family (\textit{"educate others about risks"}). Furthermore, participants reported avoiding certain actions or preventing others from doing them, e.g., to \textit{"Stop visiting high-risk areas when hazards are present"} or partaking in community events and volunteering initiatives, e.g., \textit{"Beach clean-ups"}. They also discussed shared activities, such as planning evacuation routes together with their family or making an effort to build better relationships with and take care of their neighbors (\textit{"be connected with neighbors"}). The latter was mentioned as particularly important for the elderly. Another action point concerned pets and making preparations for them. For children, participants added that they can help their families translate and understand information; \textit{"be an interpreter if kids speak a language that their parents don‘t"}.

\subsection{Card Game and Design}
Based on the survey and focus groups results and additional research about relevant local institutions, contacts, and events, we created a set of 20 action cards for our interactive tabletop game. For this, we clustered the suggestions thematically, identifying five main categories, \textit{Information Sources}, \textit{Community Engagement}, \textit{Family, Pets \& Friends}, \textit{Neighborhood Relations}, and \textit{Maintenance Tasks}, and prioritized and rephrased the actions together with the experts. See \autoref{tab:actionItems} in the appendix for a complete overview. We ignored other actions to simplify the game logic by having an even number of four actions for each of the five categories. 
For getting a high flood resilience score, players need to score points in all categories while getting the highest points possible in each category per round, and depending on the number of players. Like this, we simulated the real-life need for diverse skills and collaboration, as players can only play four cards per game but have to earn points in five categories for a high resilience score.

\paragraph{Rational} We decided to integrate the identified actions as playing cards because they provide multiple advantages: 1) they can easily be switched out and extended, 2) playing cards is a well-known and intergenerational family activity allowing a more intuitive handling, 3) they provide great flexibility in their design and 4) handing them out as little souvenirs and take-home inspirations is relatively cost efficient. The latter aspect further responds to the expert's wish to relate the game experience beyond the onsite interaction. We equipped some cards with QR codes to related websites, such as the local alert notification system, providing additional information that players could follow up on after their game interaction. In addition, we created cardholders with NFC tags for the overarching game, sending the card information to the central game logic for score updates. This also turns the cards into interactive, tangible objects, an essential component for fitting the overall game into the center's context.

\paragraph{Design} We designed the cards to support quick and easy distinction between categories, points per phase, and actions, using different background colors and icons that we will also reuse in the projection, as shown in \autoref{fig:teaser}. Further, the most important information, the points per phase, is positioned at the top and separated through circular shapes and weather-per-phase-related background images. We created the cards in Illustrator 2025 and used Adobe Firefly\footnote{\url{https://firefly.adobe.com/?media=featured}, last accessed May 27, 2025}, an online AI image generator tool, to create the scenario images representing the actions. Lastly, we aimed for bright colors and simple corporate-memphis style images to convey a positive playfulness and to motivate engagement.


\section{Discussion \& Limitations}
Identifying locally relevant actions and transferring them into game components is a complex and underresearched process~\cite{green2024,Galeote2023}. Thus, we here reflect on our process and the cards for bottom-up flood resilience.
Before, we acknowledge that our survey sample was small, leading to a limited number of actions to convert into playing cards. We do not expect our card game to be exhaustive, but it serves merely as a starting point as of now. Additionally, not all action cards are doable by everybody. We need to consider whether a player owns a house, a pet, has neighbors, uses a phone, etc., to make a judgment about the "doability" of the cards. Nonetheless, they emphasize the small levels at which everybody can contribute to increasing flood resilience for oneself, one's family, and community. 

\subsection{The Process of Creating Doable and Locally-adapted Action Cards}
Our approach comprised the observing researcher's perspective at the onsite field study, online research about local conditions, institutions, and contacts, the experts' feedback on the general situation as well as on the cards, and the online survey collecting actions done and suggested by the broader local community. Each complemented the other and was essential for identifying and defining the action cards. For one, the experts' knowledge of the local flood risks, awareness, and official countermeasures contributed to a better understanding of the current gaps and needs. Furthermore, discussing the identified actions with them supported prioritization and the identification of potential risks and mistakes. However, they had few suggestions for flood-related actions from a bottom-up perspective. The survey that we distributed across the local area and communication channels narrowed the gap with 27 actions that participants already take for themselves and others. Yet, as our small sample size also shows, engaging the local community is hard and takes a lot of effort with limited results~\cite{Rigby2023}, also because they often have no or little benefit from engaging~\cite{Hirsch2025}. Furthermore, by conducting an open-minded and naive field observation at the beginning, we allowed for a fresher outsider perspective in addition to the experts' and community's insider perspectives. Lastly, the online research supported tying the identified actions directly to local events, institutions, and places, empowering players to set game learnings into real-life actions. We will still have to test the cards' relevancy and understandability as part of the overall tabletop game. Yet, by applying complementary research methods and data, we were able to design doable and locally-adapted action cards to support bottom-up flood resilience. 

\subsection{Card Games for Bottom-up Flood Resilience}
Themed card games and collectible cards have been explored for a long time, but not in the context of flood resilience. The introduced card design and categories can be transferred and scaled to other contexts and actions, emphasizing the flexibility of action cards as game components. Identified actions differed between actions for oneself, for and with family or friends, and for and with the neighborhood community, clearly integrating it into local conditions and resources and, by that, contributing to developing games that narrow psychological and geographical distance~\cite{Galeote2024}. By gathering actions from locals and evaluating them with experts, we ensured the actions' doability for locals, empowering locals' flood resilience through self-efficacy~\cite{VINNELL2021}. In addition, each card category will contribute to receiving an overarching flood resilience score, emphasizing the multi-facted interdependencies of bottom-up community adaptations~\cite{Selje2024}. While we still have to assess the cards as part of the tabletop game, our design decisions and approach contribute to bottom-up flood resilience through an interactive card game that can further be scaled into a collectible card game by using card types, such as those developed by~\cite{Nguyen2016}. We see our learning and exploration of the action card developments transferable to other locations, natural hazard types, or card games that aim toward bottom-up community empowerment.

\section{Conclusion \& Future Work}
Our work reports and reflects on the process of identifying 27 flood resilience-related actions and turning them into 20 action cards as part of a larger tabletop game. We are at a stage of testing the cards as part of the tabletop game, and to determine how doable and locally relevant players perceive them. Furthermore, the card set as well as the tabletop game serve as a starting point for collectible cards that we plan to develop, connecting them to other community places and playful interventions related to flood resilience. For this, we aim to extend the current card set and increase collaboration with other local partners and institutions. Our work contributes an approach to identifying and transferring locally relevant and doable actions into card game components to the CHI Play community. Furthermore, our work highlights the potential of locally-adapted card games for flood resilience, which might inspire other game designers and researchers developing climate resilience games or interventions.


\begin{acks}
We thank our experts for their close collaboration. The work was supported by the UCSC Center for Coastal Climate Resilience, CA, USA.
\end{acks}

\bibliographystyle{ACM-Reference-Format}
\bibliography{library,JamesReferences}

\pagebreak
\section{Attachment}

\subsection{Action list} \label{Attachment}

\begin{table*} [!]
\centering
\caption{Action categories, individual actions, and scores as integrated into the game. The numbers I-IV describe the scores per phase, I: long before, and IV: after a flood. }
\def\arraystretch{1.2}
\begin{tabular}{ c|m{10cm} cccc} 
\hline
   \textbf{Category} & \textbf{Action} & \textbf{I} & \textbf{II} & \textbf{III} & \textbf{IV} \\
\toprule  
\multirow{5}{*}{\rotatebox[origin=c]{90}{\parbox{1.2cm}{Community Engagement}}}
    & You participated in a workshop about flood preparation and recovery, such as CERT. & 4 & 2 & 0 & 3 \\
    & You joined one of the volunteering days to secure the river bed's ecosystem. & 4 & 2 & 1 & 3\\
    & You donate supplies or money to people or institutions like the Santa Cruz Community Foundation. & 2 & 4 & 4 & 4\\
    & You work with a school to create a resource map to identify shelters, supply hubs, or hidden dangers in your neighborhood. & 3 & 4 & 4 & 3 \\
    \midrule
    \multirow{6}{*}{\rotatebox[origin=c]{90}{\parbox{1.2cm}{Information Sources}}}
   &You subscribe to the county's flooding and rainfall news on CruzAlert, X or Facebook. & 3& 3& 2&1\\
  & You learn about your home's level of flood risk by consulting local maps and learning about your watershed. &4&3&1&1\\
   &A flood warning came in for your area, and you informed your neighbors online and on your block about it. &0&3&4&0\\
  & You report downed trees/blocked roads on your community's X. &2&2&4&4\\
      \midrule
    \multirow{6}{*}{\rotatebox[origin=c]{90}{\parbox{1.2cm}{Maintenance Tasks}}}
 & You cleaned the drainpipes and gutters of your house.  &1&3&3&1\\
  & With your neighbors, you clean the street in front of your house. &2&4&3&2\\
  & You go for the regular bicycle checkups, so you will not become a safety hazard in bad weather conditions. &2&2&0&1\\
 & The city needs help in checking and maintaining drains across the town. You volunteer half a day and covered a large area.  &2&4&3&4\\
     \midrule
    \multirow{7}{*}{\rotatebox[origin=c]{90}{\parbox{1.2cm}{Neighborhood Relations}}}
 & You know where your family's generator is and switch it on for your own house and your neighbor's.  &1&3&4&2\\
  & You go check on your neighbors and see if they need any help getting to safety. &4&2&1&2\\
  & You introduced yourself to your neighbors and exchanged phone numbers for emergency cases. &4&2&1&2\\
  & You and your neighbors install plants to create permeable surfaces. &4&2&0&1\\
      \midrule
    \multirow{8}{*}{\rotatebox[origin=c]{90}{\parbox{1.2cm}{Family, Pets and Friends}}}
 & You help your family to understand the latest evacuation instructions from official authorities.  &3&4&4&2\\
 & Your friends want to make clips of the dangerous waves that happen along the coast. You stop them from going there.  &1&3&4&2\\
  & You prepare your pet's safety pack to keep it safe in the next emergency, following ASPCA recommendations. &4&4&2&2\\
 &  Your family fosters a pet while another family gets their house restored. &1&1&4&4\\
 \bottomrule
\end{tabular}
\label{tab:actionItems}
\end{table*}

\subsection{Attachment: Qualtrics Survey} \label{qualtricsSurvey}
Here is an excerpt of our online survey. Block one included the study information and consent form, which we excluded.
\includepdf[pages=-]{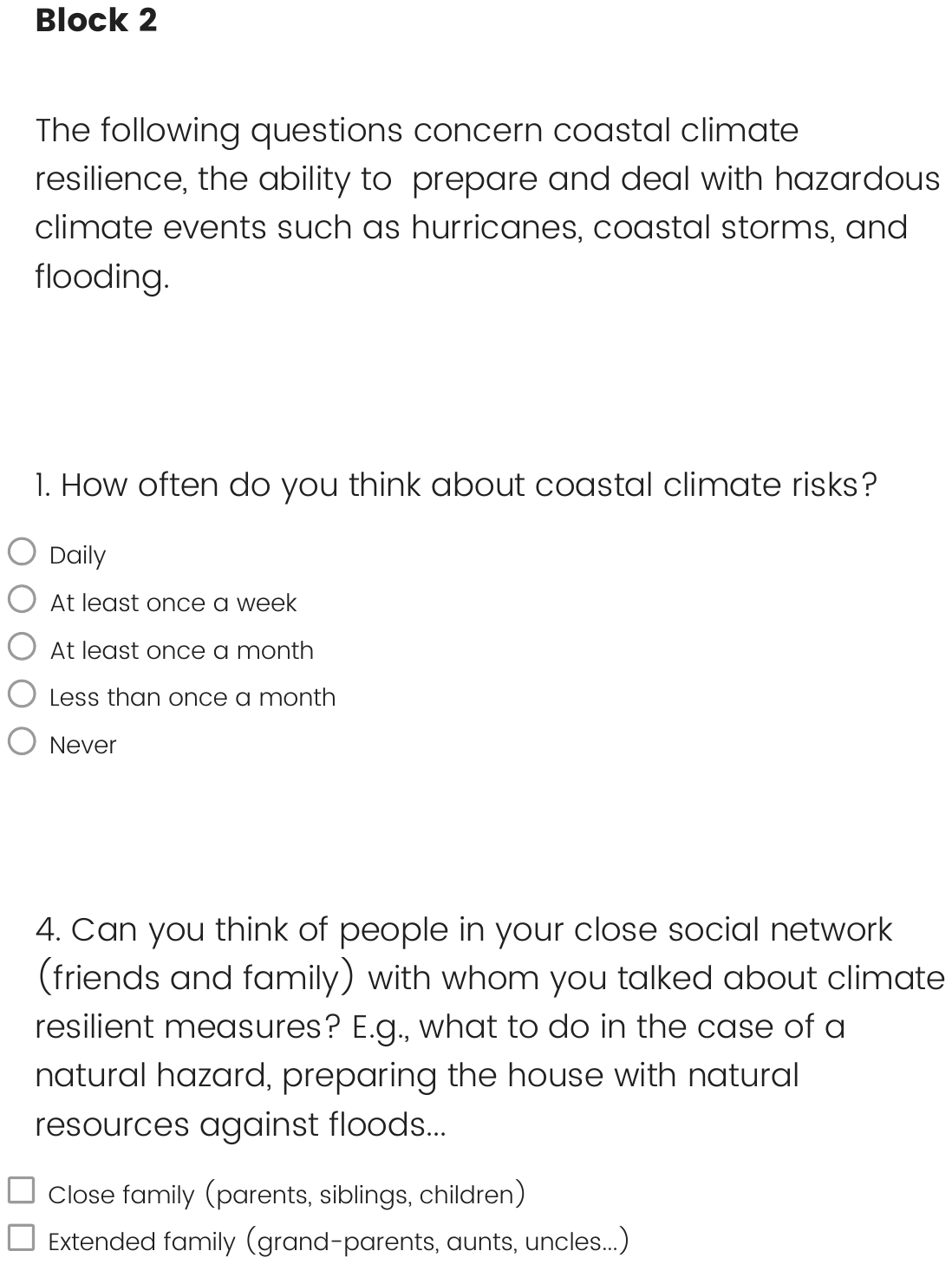}

\end{document}